\def\tipo{2}

\ifnum \tipo = 2
 \documentstyle[prl,twocolumn,aps,psfig]{revtex}
 \def\figsize{8.5cm}
 \def \frontmatter{\twocolumn[\hsize\textwidth\columnwidth\hsize\csname@twocolumnfalse\endcsname}
 
\else
 \documentstyle[preprint,aps,prl,psfig]{revtex}
 \def\figsize{12cm}
 \def\frontmatter{}
 
\fi

\begin{document}
\draft
\frontmatter

\title{Correlation Differences in Heartbeat Fluctuations During Rest and Exercise} 

\author{Roman Karasik$^1$, Nir Sapir$^1$, Yosef Ashkenazy$^2$, Plamen Ch. Ivanov$^{3,4}$,\\
Itzhak Dvir$^5$, Peretz Lavie$^6$, and Shlomo Havlin$^1$}

\address{
$^1$ Department of Physics and 
Gonda-Goldschmied-Center for Medical Diagnosis,\\
Bar-Ilan University, Ramat-Gan 52900, Israel\\
$^2$ Center for Global Change Science, Massachusetts Institute of Technology,\\
MIT Room 54-1726, Cambridge, MA 02139, USA\\
$^3$ Center for Polymer Studies and Department of Physics, Boston University,
Boston, MA 02215, USA\\
$^4$ Harvard Medical School, Beth Israel Deaconess Medical Center, 
Boston, MA 02215, USA\\
$^5$ Itamar Medical Ltd. Cesarea, Israel\\
$^6$ Sleep Laboratory, Faculty of Medicine, Technion-Israel Institute
of Technology, Haifa, Israel}
\date{\today}
\maketitle

\begin{abstract}
We study the heartbeat activity of healthy
individuals at rest and during exercise. We focus on correlation
properties of the intervals formed by successive peaks in the pulse wave  
and find significant scaling differences
between rest and exercise.   
For exercise the interval series is anticorrelated at short time scales
and correlated at intermediate time scales, while for rest we observe the
opposite crossover pattern --- from strong correlations in the short-time regime
to weaker correlations at larger scales. We suggest a physiologically motivated 
stochastic scenario to explain the scaling differences between rest and exercise and 
the observed crossover patterns.
\end{abstract}
\ifnum \tipo = 2
]
\fi

\def\figureI{
\begin{figure}[thb]
\centerline{\psfig{figure=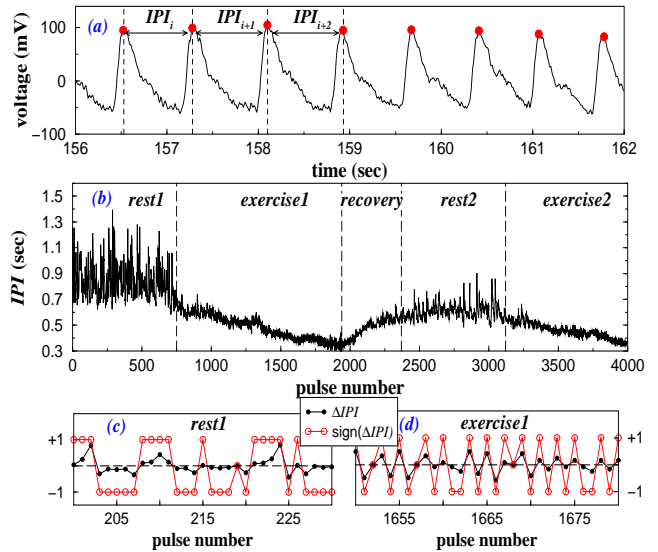,width=\figsize,height=7.5cm,angle=-90}}
{
\ifnum\tipo=2
\vspace*{0.0truecm}
\fi
\caption{\label{Fig.1} 
(a) A typical example of a pulse wave measured as a function of time.
As in the case of the electrocardiogram signal where interbeat interval fluctuations 
are studied (see, e.g., \protect\cite{peng95}), we analyze the interpulse 
intervals ({\it IPI}) between successive peaks in the pulse wave. 
(b) {\it IPI} series obtained from the pulse wave signal shown in (a). 
Each record includes 
two rest and two exercise stages. The duration of each stage varies from subject to 
subject and is between 6-10 minutes.  
(c) Sign series obtained from the increments $\Delta IPI$ in the interpulse intervals
during rest and (d) during exercise. 
Note that the sign series of the exercise
regime exhibits more frequent alternations (stronger anticorrelated behavior)
compared to rest. 
}}
\end{figure}
}

\def\figureII{
\begin{figure}[thb]
\centerline{\psfig{figure=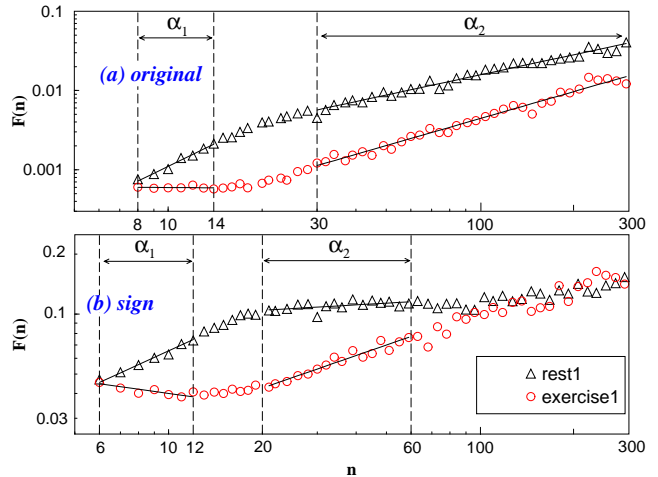,width=\figsize,height=6.5cm,angle=-90}}
{
\ifnum\tipo=2
\vspace*{0.0truecm}
\fi
\caption{\label{Fig.2} 
Fluctuation function,
{\it F(n)}, as a function 
of time scale {\it n} (in beat number) for rest~($\bigtriangleup$) and exercise~($\circ$) 
stages of a typical healthy subject for 
(a) the original {\it IPI} series, and (b) the sign series $sign(\Delta IPI)$ . 
For all records we observe a crossover between two different regimes
of correlations. The dashed lines indicate the boundaries of these regimes in which 
short-range scaling exponents $\alpha_{1}$ and intermediate exponents $\alpha_{2}$ have 
been calculated. 
Note the different crossover patterns for rest ($\alpha_{1}>\alpha_{2}$) 
and exercise ($\alpha_{1}<\alpha_{2}$) stages. 
}}
\end{figure}
}

\def\figureIII{
\begin{figure}[thb]
\centerline{\psfig{figure=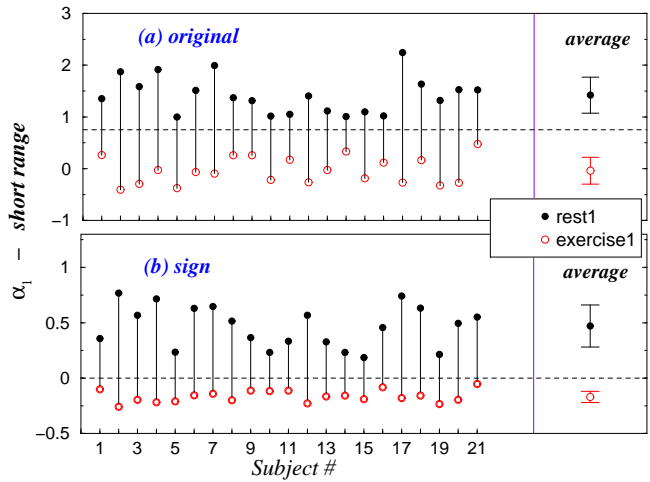,width=\figsize,height=6.5cm,angle=-90}}
{
\ifnum\tipo=2
\vspace*{0.0truecm}
\fi
\caption{\label{Fig.3} 
Short-range scaling exponents $ \alpha _{1}  $
from 21 healthy individuals (a) for the original {\it IPI} series and (b) for the sign 
series at rest~$(\bullet)$ and  during exercise~$(\circ)$. 
At the right hand side we show the average $\alpha_{1}$ $\pm$ standard
deviation. In accordance with Fig.~\ref{Fig.2}, the short-range exponents during 
exercise are significantly smaller than during rest. Note the complete separation 
of the two stages, emphasized by the dashed lines. 
}}
\end{figure}
}

\def\tableI{
\vskip 0.15in
\begin{tabular}{|c|c|c|c|c|}
\hline
 &
\multicolumn{2}{|c|}{original {\it IPI} series} &
\multicolumn{2}{|c|}{$sign(\Delta IPI)$}
	\\ \cline{2-5} 
& $\alpha _{1}$ & $\alpha _{2}$ & $\alpha _{1}$ & $\alpha _{2}$ \\ \hline \hline
{\it rest1} & {\small 1.42} $\pm$ {\small 0.35} & {\small 0.78} $\pm$ {\small 0.16} &
{\small 0.47} $\pm$ {\small 0.19} & {\small 0.17} $\pm$ {\small 0.11} \\ \hline 
{\it rest2} & {\small 1.43} $\pm$ {\small 0.30} & {\small 0.75} $\pm$ {\small 0.17} &
{\small 0.38} $\pm$ {\small 0.26} & {\small 0.15} $\pm$ {\small 0.12}\\ \hline \hline
{\it ex1} & {\small -0.04} $\pm$ {\small 0.26} & {\small 1.07} $\pm$ {\small 0.18} &
{\small -0.17} $\pm$ {\small 0.05} & {\small 0.41} $\pm$ {\small 0.09} \\ \hline
{\it ex2} & {\small -0.14} $\pm$ {\small 0.17} & {\small 1.11} $\pm$ {\small 0.16} &
{\small -0.21} $\pm$ {\small 0.06} & {\small 0.41} $\pm$ {\small 0.1} \\ \hline \hline
{\it whole} & {\small 1.21} $\pm$ {\small 0.25} & {\small 0.91} $\pm$ {\small 0.12} &
{\small 0.23} $\pm$ {\small 0.15} & {\small 0.22} $\pm$ {\small 0.07} \\ \hline
\end{tabular}

\begin{table}
\caption{\label{Table1}
Comparison of scaling exponents $\alpha_{1}$ and $\alpha_{2}$ between rest stages, 
exercise stages and whole records that include rest and exercise episodes altogether. 
For each stage the average scaling exponent $\pm$ standard deviation are shown.
}
\end{table}
}

\def\figureIV{
\begin{figure}[thb]
\centerline{\psfig{figure=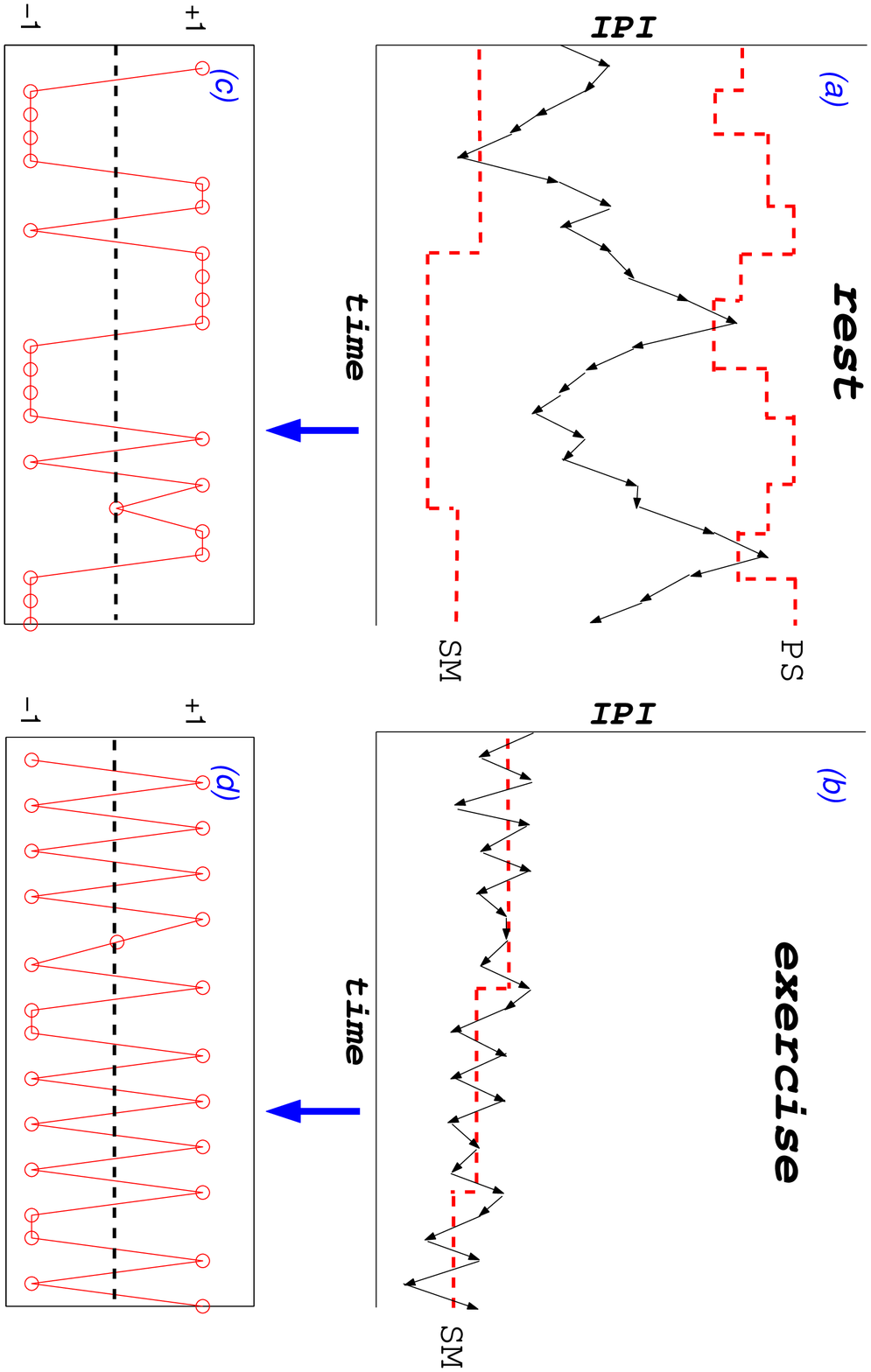,width=\figsize,height=6cm,angle=90}}
{
\ifnum\tipo=2
\vspace*{0.0truecm}
\fi
\vskip 0.1in
\caption{\label{Fig.4} 
Schematic illustration of a random walk with (a) two 
different levels of attraction and (b) a single attraction level, and their sign
decomposition (c) and (d). (a) The combined effect of the upper level (representing
the ``preferred'' level of parasympathetic (PS) system) and the lower level (sympathetic 
(SM) system) do not restrict the walker's fluctuations for the short-time regime. 
But for the larger times the bounded walk results in a crossover to a less correlated
behavior, similar to the crossover obtained for the rest stage (Fig.~\ref{Fig.2}).
(b) On the other hand, the walker attracted by a single SM level produces  
short-range anticorrelations with a crossover to a more correlated behavior in the 
intermediate regime due to the alterations of the SM level (mimicking exercise). 
The sign series for the two attractive levels presented in (c) is more likely to cluster than 
the series for a single level scenario presented in (d) (Compare with sign dynamics for rest 
and exercise shown in Fig.~\ref{Fig.1}c and  Fig.~\ref{Fig.1}d).
}}
\end{figure}
}

One of the important questions in the analysis of complex physiological time
series is how such series reflect the dynamical properties
associated with the underlying control mechanism~\cite{liebovitch,shlesinger}.
Recently, it was found, e.g., that the fluctuations of the
heart interbeat intervals reveal long-range power-law correlations~\cite{peng93} 
and hidden scale-invariant structure~\cite{ivanov96} which may
be useful for diagnosis and prognosis~\cite{huikuri2000}. 
Here we study the correlation (scaling) properties of heartbeat dynamics as 
reflected by the pulse wave measured from the finger~\cite{itamar,schnall99}.

Previous studies of long interbeat interval series have focused primarily on 
24h~\cite{peng95} and 6h~\cite{ivanov99} records, which include 
periods of rest as well as periods of a more intensive physical 
activity. However, heartbeat dynamics can change dramatically with physical activity.
Thus important differences in cardiac regulation associated with 
rest and exercise may not be clearly 
seen when analyzing records which mix together rest and exercise regimes. 
Here we consider rest and exercise activities {\it  separately}. We focus on the 
correlations in the interpulse interval 
({\it IPI}) series derived from the pulse wave signal during rest and 
exercise (Fig.~\ref{Fig.1}a).
By studying the changes in the correlation properties we wish to 
achieve a better understanding of the physiological mechanism that 
regulates heartbeat dynamics at rest and during physical exercise.

We analyze 21 records from healthy subjects. Each record includes 4 different 
stages of physical activity denoted as {\it rest1}, {\it exercise1},
{\it rest2} and {\it exercise2} (Fig.~\ref{Fig.1}b). At the first
stage ({\it rest1}) we measure the {\it IPI} under normal rest conditions. At
the next stage ({\it exercise1}) subjects are asked to run on a treadmill. 
After a short {\it recovery}, during which subjects sit down to recover their 
heart rate, a new rest-exercise episode (denoted as {\it rest2} and
{\it exercise2}) is followed.

\figureI

To study the correlation properties of the {\it IPI} series during rest and exercise 
stages we use the detrended fluctuation analysis (DFA)~\cite{peng94} which is a 
method developed to avoid
spurious detection of correlations that are artifacts of trends related to 
nonstationarity. 
The DFA procedure consists of the following
steps. We first integrate the {\it IPI} series to construct the profile 
$Y(k)=\sum^{k}_{i=1}(IPI_{i}-\left\langle IPI\right\rangle)$ 
where $ \left\langle IPI \right\rangle  $ is the series average. Next,
we divide the integrated signal, $Y(k)$, into equal windows of size
{\it n} and find the local trend in each window by a least-squares
polynomial fit. The order of the polynomial fit specifies the order of
the DFA~\cite{remark1,bunde2000}. Then we calculate the average of the square distances 
around the local trend. This procedure is repeated to obtain the root mean square
fluctuation function $F(n)$ for different window sizes $n$. 
A power-law relation between $F(n)$ and $n$, $F(n)\sim n^{\alpha}$, indicates 
the presence of scaling in the series.
According to random walk theory, the scaling exponent $\alpha$ is related to the 
autocorrelation function exponent $\gamma$ ($C(n)\sim n^{-\gamma }$ when $0<\gamma<1$) 
and to the power spectrum exponent $\beta$ ($S(f)\sim 1/f^\beta$) 
by $\alpha=1-\gamma/2=(\beta+1)/2$~\cite{vicsek}.
The value $\alpha =0.5$ indicates that 
there are no (or finite-range) correlations in the data.
When $\alpha <0.5$ the signal is {\it  anti-correlated}, meaning that large
values are most probable to be followed by small values. The case of $\alpha >0.5$
indicates the existence of persistent behavior in the time series,
meaning that large values are most probable to be followed by large values.
The higher $\alpha$ is, the stronger are the correlations in the signal. 

We also study the correlation properties of the sign series  
$sign(\Delta IPI)$~\cite{ashkenazy2000}, derived from the {\it IPI} 
increments $\Delta IPI_i=IPI_{i+1}-IPI_{i}$~\cite{remark2}.
Fig.~\ref{Fig.1}c and Fig.~\ref{Fig.1}d show representative examples
of sign series obtained from rest and exercise stages
respectively. For exercise, the signs of {\it IPI} increments tend to
alternate rapidly, indicating a strong anticorrelated behavior. At
rest stage, on the other hand, the signs alternate every several
points, and thus this dynamics may be characterized by a more correlated behavior. 

Due to the fact that during exercise the {\it IPI} series exhibits strong short-range 
anticorrelations, we first integrate the {\it IPI} series for all rest and exercise 
episodes (in addition to the integration built in the DFA method), to avoid inaccurate 
estimation of the scaling exponents for exercise segments.
Because of the apparent linear decrease of the {\it IPI} during the 
exercise stage (Fig.~\ref{Fig.1}b), the extra integration introduces a 
parabolic trend. To eliminate the effect of this parabolic trend in the exercise stage, 
we perform $3^{\rm rd}$ order DFA~\cite{remark3} on the {\it integrated} {\it IPI} series. 
The integration procedure is not necessary for evaluating $\alpha$ for the rest
episodes, since they exhibit correlated behavior~\cite{remark4}. 

In Fig.~\ref{Fig.2}a we present the fluctuation function, $F(n)$, of the integrated 
{\it IPI} series~\cite{remark5} for rest and exercise segments of a typical subject.     
For all 21 individuals we observe a characteristic
crossover around $n\approx 20$ where there is a change in the correlation
behavior between short and intermediate scales regimes. We denote the
scaling exponent of the short-range regime as $\alpha _{1}$
(estimated for scales $ 8\leq n\leq 14$) and the scaling exponent
of the intermediate regime as $\alpha _{2}$ (estimated for $30\leq n\leq 300$). 
The type of crossover is different for
rest and exercise: for the rest 
$\alpha _{1} > \alpha _{2}$, while for the exercise 
$\alpha _{1} < \alpha _{2}$. The fluctuation functions for rest and exercise
stages construct a ``fish''-like curve (Fig.~\ref{Fig.2}). 

\figureII

We apply a similar scaling analysis to sign series
derived from rest and exercise segments of the {\it IPI} signal
(Fig.~\ref{Fig.2}b). 
The sign series do not have any global trend, like the original {\it IPI} series have; 
thus, in this case it is enough to use $2^{\rm nd}$ order DFA. For sign series we 
calculate the short-range scaling exponent $\alpha _{1}$ in the range $ 6\leq n\leq 12$
and intermediate exponent $\alpha _{2}$ in the range $ 20\leq n\leq 60$~\cite{remark6}.  

\tableI

We obtain a complete separation between rest and exercise for the original and sign 
series in the short range (Fig.~\ref{Fig.3}) and an almost
complete separation for the intermediate-range scaling exponents
(Table~\ref{Table1}). The {\it p-values} for the original and sign series 
(obtained by the paired samples Student's {\it t-test} \cite{NR}) 
are less than $10^{-10}$ for the short-range regime and 
less than $10^{-4}$ for the intermediate regime. We find that our results 
are robust and do not change significantly with repetitive rest and exercise stages
(Table I). 

\figureIII

To illustrate the importance of considering separately rest and exercise episodes, 
we perform our analysis also on the entire {\it IPI} records which include rest and 
exercise episodes altogether (Fig.~\ref{Fig.1}b). We find indeed that the scaling
of the whole record reflects neither the correlation properties of rest, nor
of exercise (see Table~\ref{Table1}).
   
The significant differences between the values of $\alpha_{1}$ (Fig.~\ref{Fig.3}) and
the different crossover patterns for rest and exercise stages (Fig.~\ref{Fig.2}) may offer
some insight on the underlying  physiological mechanism controlling the heartbeat 
dynamics. 
The heart rhythm is regulated mainly by the parasympathetic (PS) and the 
sympathetic (SM) branches of the autonomic nervous systems~\cite{berne}.
PS impulses slow the heart rate while 
SM impulses accelerate it. The interaction between these two branches is 
reflected by the time organization of the {\it IPI} series (Fig.~\ref{Fig.1}b).

The principle of homeostasis (dynamic equilibrium)
asserts that physiological systems seek to maintain a constant output
in spite of continuous perturbations~\cite{bernard}. 
However, healthy systems even at rest
display highly irregular dynamics (see Fig.1b)~\cite{liebovitch,shlesinger,akselrod81}. 
In a recent work
Ivanov et al.~\cite{ivanov98} proposed a general approach based on the concept of
stochastic feedback to account for the complex fractal variability in
biological rhythms. In this framework the time evolution of a
physiologic system, e.g. the heartbeat dynamics, can be represented by
a random walk biased toward some preferred ``attracting'' levels. Both
the SM and PS systems controlling the heart rhythm
generate attracting levels which bias the walker (modelling the interbeat interval
series) in opposite directions leading to complex heartrate
fluctuations. Although these attracting levels change in time, 
according to the response of the intrinsic physiological mechanism, 
they can vary in a limited range only, thus keeping the walker away 
from extreme values.

\figureIV

Based on this general approach we suggest a schematic scenario that explains the
different crossover patterns of the {\it IPI} fluctuations for rest and 
exercise (Fig.~\ref{Fig.2}). 
At rest both the SM and PS systems are active,
and each of them attracts the walker toward its own level (Fig.4a). Since
the response time of the PS system is shorter than that
of the SM system~\cite{warner62,ivanov98}, we assume that the
preferred attracting level of the PS system alters more
rapidly that the one related to the SM system. This scenario
can account for the crossover observed for the rest stage
(Fig.2). When the walker is between the two attracting levels, each
level imposes a bias in an opposite direction. Thus the walker is able
to move in both directions until he crosses any of the two levels
after which he is pulled back.  This model scheme reproduces the
crossover in the scaling behavior (Fig.2) from a larger value of the
correlation exponent at short scales, where the fluctuations of the
walker are not bounded, to a lower value of the exponent at large time
scales, where the dynamics of the walker is bounded by the SM
and PS attracting levels.

During exercise the SM system dominates~\cite{robinson66} and the dynamics can 
be described effectively by a single attracting level (Fig.4b).  In this case the
walker fluctuates around this level producing an anticorrelated behavior at
short time scales. However, since the attracting level changes with time
and since the walker follows
these changes, the fluctuations in the random walk increase at intermediate
time scales, causing a crossover to a more correlated behaviour. This scheme
accounts for the observed crossover pattern in the scaling of the IPI
fluctuations from an anticorrelated behavior with small value of the
corretalion exponent at short time scales to a correlated behavior
characterized by  a larger value of the exponent at large time scales (Fig.2).
Our scenario can  also explain the remarkable difference in the amplitude of
the fluctuations at rest and during exercise (see Fig.1b). When two attracting
levels bias the walker (a situation in our scenario corresponding to rest)
the fluctuations are larger compared to the exercise stage when there
is a single dominant attracting level (Fig.4).

A ``fish'' structure similar to Fig.~\ref{Fig.2} (but with different scaling exponents) 
was observed when comparing 
healthy subjects with congestive heart failure patients~\cite{peng95}.
These results support our scenario, since for heart failure 
patients there are evidences of SM dominance~\cite{saul88}, resembling the state of 
the autonomic nervous system under physical exercise.
  
In summary, we study correlations in heartbeat fluctuations during rest
and exercise.
We show that the significant scaling differences and the different crossover 
patterns between rest and exercise (Fig.~\ref{Fig.2}) can be explained based on 
the ``attractive levels'' scenario. We, therefore, conclude that the interaction 
between the competing branches of the autonomic nervous system underlies the 
correlation properties of heartbeat.

We wish to thank J.W. Kantelhardt and J.M. Hausdorff for helpful discussions. 
This work was supported by the Binational Israel-USA Science Foundation and the 
NIH/National Center for Research Resources (P41 RR13622).


\begin{references}
\bibitem{remark2} If, due to finite resolution, $\Delta IPI_{i}=0$, we
define $sign(\Delta IPI_{i})=0$.
\bibitem{remark3}
We found that scaling exponents calculated using $4^{th}$ order DFA are very close 
to the exponents obtained by $3^{rd}$ order DFA. 
This indicates that we have reached the ``real'' scaling which is not an artifact of the 
global trend during exercise.
\end{references}
\end{document}